\documentclass{aa}
\usepackage{epsfig}

\begin{document}

   \thesaurus{}
   \title{$\theta$ Hya: 
Spectroscopic identification of a second B star$+$white dwarf binary}

   \author{M. R. Burleigh \and M. A. Barstow}

   \offprints{Matt Burleigh, mbu@star.le.ac.uk}

   \institute{Department of Physics and Astronomy, University of
                Leicester,
                Leicester LE1 7RH, UK }

   \date{Received 	; accepted 	}

    \titlerunning{$\theta$ Hya: a second B star$+$white dwarf binary} 

    \authorrunning{M. R. Burleigh \and M. A. Barstow }

\maketitle

\begin{abstract}

We report the identification, in an Extreme Ultraviolet Explorer (EUVE)
spectrum, of a hot white dwarf companion to the 3rd magnitude late-B star
$\theta$ Hya (HR3665, HD79469). This is the second B star$+$white dwarf
binary to be conclusively identified; Vennes, Bergh\"ofer and 
Christian (1997), and Burleigh and Barstow (1998) had previously 
reported the spectroscopic discovery of a hot white dwarf companion to 
the B5V star y Pup (HR2875). 
Since these two degenerate stars must have evolved from
main sequence progenitors more massive than their B star companions, they
can be used to place observational lower limits on the maximum mass for
white dwarf progenitors, and to investigate the upper end of the
initial-final mass relation. Assuming a pure hydrogen composition, we
constrain the temperature of the white dwarf companion to $\theta$ Hya 
to lie between 25,000K and 31,000K. We also predict that a third bright
B star, 16 Dra (B9.5V), might also be hiding an unresolved hot white
dwarf companion.

   \end{abstract}

%
%  14.Sep.'90: Demo-Vs.
%________________________________________________________________

\section{Introduction}

Prior to the extreme ultraviolet (EUV) surveys of the ROSAT Wide Field
Camera (WFC, Pye et~al. 1995) and NASA's Extreme Ultraviolet Explorer
(EUVE, Bowyer et~al. 1996), only a handful of binary systems consisting
of a normal star (spectral type K5 or earlier) 
plus a degenerate white dwarf had been identified. Some
of these systems, like the prototype Sirius (A1V$+$DA), are relatively
nearby and wide enough that the white dwarf can be readily resolved
from its bright companion. Most of these types of binary, however, are all but
unidentifiable optically since the normal stellar companion completely
swamps the flux coming from the white dwarf. The detection
by ROSAT and EUVE 
of EUV radiation with the spectral signature of a hot white dwarf originating
from apparently normal, bright main sequence stars, therefore, 
gave a clue to the
existence of a previously unidentified population of Sirius-type
binaries, and around 20 new systems have now been identified (e.g.
Barstow et~al. 1994, Burleigh, Barstow and Fleming 1997, Burleigh 1998
and  Vennes, Christian and Thorstensen 1998, hereafter VCT98). In each case, 
far-ultraviolet spectra taken with the
International Ultraviolet Explorer (IUE) were used to confirm the
identifications. This technique proved excellent for finding systems
where the normal star is of spectral type $\sim$A5 or later, since the
hot white dwarf is actually the brighter component in this wavelength
regime ($\sim$1200$-$$\sim$2000{\AA}).    
Unfortunately, even at far-UV wavelengths, stars of spectral types 
early-A, B and O will completely dominate any emission from smaller, fainter,
unresolved companions, rendering them invisible to IUE. 

$\theta$
Hya (HR3665, HD79469) and y Pup (HR2875) are two bright B stars
unexpectedly detected in the ROSAT and EUVE surveys. Their soft X-ray and
EUV colours are similar to known hot white dwarfs, so it was suspected
that, like several other bright normal stars in the EUV catalogues, they
were hiding hot white dwarf companions. However, for these two systems 
it was, of course,  
not possible to use IUE or HST to make a positive identification,
and instead we had to wait for EUVE's spectrometers to make a pointed
observation of each star. y Pup was observed in 1996, and the formal
discovery of its hot white dwarf companion was reported by Vennes,
Bergh\"ofer and Christian (1997), and Burleigh and Barstow (1998). $\theta$
Hya was observed by EUVE in February 1998 and the EUV continuum
distinctive of a hot white dwarf was detected (Figure 1). This spectrum
is presented, analysed and discussed in this letter.

\begin{table*}
\begin{center}
\caption{X-ray and EUV count rates (counts/ksec)}
\begin{tabular}{llcccccccc}
 &  & WFC & & PSPC & & EUVE & & & \\
ROSAT No. & Name & S1 & S2 & (0.1-0.4keV) & (0.4-2.4keV) & 100\AA
& 200\AA & 400\AA & 600\AA \\
RE J0914$+$023 & $\theta$ Hya & 52$\pm$7 & 148$\pm$12 & 124$\pm$24 & 0.0 &
122$\pm$15 & 0.0 & 0.0 & 0.0 \\
\end {tabular}
\end{center}
\vspace{-0.5cm}
\end{table*}

White dwarf companions to B stars are of
significant importance since they must have evolved from massive
progenitors, perhaps close to the maximum mass for white dwarf progenitor
stars, and they are likely themselves to be much more massive than the
mean for white dwarfs in general (0.57M$_\odot$, Finley, Koester and
Basri 1997). The value of the maximum mass feasible for producing a
white dwarf is a long-standing astrophysical problem. Weidemann (1987)
gives the upper limit as 8M$_\odot$ in his semi-empirical initial-final
mass relation. Observationally, the limit is best set by 
the white dwarf companion to y Pup, which must have
evolved from a progenitor more massive than B5 (6$-$6.5M$_\odot$).

\section{The main sequence star $\theta$ Hya} 

$\theta$ Hya is a V$=$3.88 high proper motion star; {\it
Hipparcos} measures the proper motion components as
112.57$\pm$1.41 and $-$306.07$\pm$1.20 milli-arcsecs. per year. 
$\theta$ Hya was originally classified as a
$\lambda$ Boo chemically peculiar star, although from
ultraviolet spectroscopy Faraggiani, Gerbaldi and B\"ohm 
(1990) later concluded that
$\theta$ Hya was not in fact chemically peculiar, a finding backed up by
Leone and Catanzaro (1998). Their derived abundances from high resolution
optical spectroscopy are almost coincident with expected main sequence
abundances. The
{\it SIMBAD} database, Morgan, Harris and Johnson (1953) and 
Cowley et~al. (1969) give the spectral type 
as B9.5V. VCT98 note that it is a fast rotator
(v$_{rot}$sin i$\sim$100 km s$^{-1}$), and that the detection of HeI at 
4471{\AA} also suggests a B star classification. 

\section{Detection of EUV radiation from $\theta$~Hya 
in the ROSAT WFC and EUVE surveys}

The ROSAT EUV and X-ray all-sky surveys were conducted between July 1990
and January 1991; the mission and instruments are described elsewhere
(e.g. Tr\"umper 1992, Sims et~al. 1990). $\theta$ Hya is associated with
the relatively bright WFC source RE J0914$+$023. The same EUV source was
later detected in the EUVE all-sky survey (conducted between July 1992
and January 1993). This source is also coincident with a ROSAT PSPC soft
X-ray detection. The count rates from all three instruments are given in
Table~1. The WFC count rates are taken from the revised 2RE Catalogue (Pye
et.~al. 1995), which was constructed using improved methods for source
detection and background screening. The EUVE count rates are
taken from the revised Second EUVE Source Catalog (Bowyer et~al. 1996).
The PSPC count rate was obtained via the World Wide Web from the on-line
ROSAT All Sky Survey Bright Source Catalogue maintained by the Max Planck
Institute in Germany (Voges et~al. 1996)
\footnote
{http://www.rosat.mpe-garching.mpg.de/survey/rass-bsc/cat.html}. 
As with y Pup (RE J0729$-$388), the EUV and soft X-ray colours and count
rate ratios are similar to known hot white dwarfs. The EUV radiation is
too strong for it to be the result of UV leakage into the detectors (see
the discussion in Burleigh and Barstow 1998). $\theta$ Hya is also
only seen in the soft 0.1$-$0.4 kev PSPC band; only one (rather unusual)
white dwarf has ever been detected at higher energies (KPD0005$+$5105,
Fleming, Werner and Barstow 1993), while most active stars are also hard X-ray
sources. Indeed, in a survey to find OB-type stars in the ROSAT X-ray
catalogue by Bergh\"ofer, Schmitt and Cassinelli (1996), only three of
the detected B
stars are not hard X-ray sources: y Pup (confirmed B5V$+$white dwarf),
$\theta$ Hya and 16 Dra (B9.5V, see
later). Therefore, Burleigh, Barstow and Fleming (1997) and VCT98   
suggested that $\theta$ Hya, like y Pup and nearly twenty
other bright, apparently normal stars in the EUV catalogues, 
might be hiding a hot white dwarf companion.

\section{EUVE pointed observation and data reduction}

\begin{figure*}
\vspace*{7cm}
\includegraphics{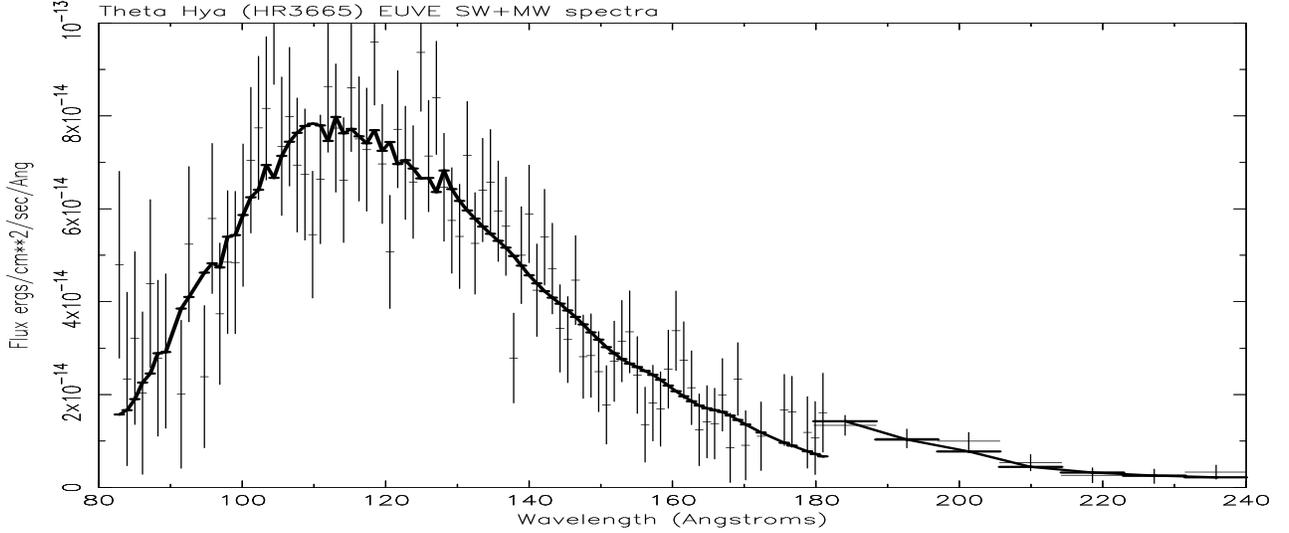}
\caption{\em EUVE short wavelength spectrum of $\theta$ Hya. 
Also shown is a pure
hydrogen white dwarf $+$ ISM model for log $g=$8.5, $T_{eff}=$28,500K, 
$N_{HI}$$=$6.6$\times$10$^{18}$atoms cm$^2$,
$N_{HeI}$$=$7.4$\times$10$^{17}$atoms cm$^2$, and
$N_{HeII}$$=$2.8$\times$10$^{17}$atoms cm$^2$. }
\end{figure*}

$\theta$ Hya was observed by EUVE in dither mode in four separate
observations in 1998 February/March for a
total exposure time of $\approx$210,000 secs. We have extracted the
spectra from the images ourselves using standard IRAF procedures. Our
general reduction techniques have been described in earlier work (e.g.
Barstow et~al. 1997). 

The target was detected in both the short (70$-$190{\AA}) and medium
(140$-$380{\AA}) wavelength spectrometers (albeit weakly), 
but not the long wavelength
(280$-$760{\AA}) spectrometer. To improve the signal/noise, we have
co-added the four separate observations, binned the short wavelength
data by a factor four, and the medium wavelength data by a factor 16. The
resultant spectrum, shown in Figure 2, reveals the now familiar EUV
continuum expected from a hot white dwarf in this spectral region.

The only stars other than white dwarfs whose photospheric EUV radiation
has been detected by the ROSAT WFC and EUVE are the bright B giants $\beta$ CMa
(B1II$-$III, Cassinelli et~al. 1996) and $\epsilon$ CMa (B2II, Cohen
et~al. 1996). The photospheric continuum of $\epsilon$ CMa is visible
down to $\sim$300{\AA}, although no continuum flux from $\beta$ CMa is
visible below the HeI edge at 504{\AA}. 
Both stars also have strong EUV and X-ray emitting winds, and in
$\epsilon$ CMa emission lines are seen in the short and medium wavelength
spectrometers from e.g. high ionisation species of iron. Similarly,
strong narrow emission features of e.g. oxygen, nickel and calcium  
are commonly seen in EUV spectra of active
stars and RS CVn systems. Since no such features are visible in the
$\theta$~Hya EUVE spectrum, we can categorically 
rule out a hot wind or a hidden active
late-type companion to $\theta$~Hya as
an alternative source of the EUV radiation.

\section{Analysis of the hot white dwarf}

\begin{table}
\begin{center}
\caption{Hamada-Salpeter zero-temperature mass-radius relation}
\begin{tabular}{ccccc}
log $g$ & $M_{WD}$ & $R_{WD}$ & $R_{WD}$ & ($R_{
WD}$/$D$)$^2$ \\
 & M$_\odot$ & R$_\odot$ & $\times$10$^6$m & {\em where D$=$39.5pc} \\
7.5 & 0.30 & 0.017 & 11.832 & 9.4243$\times$10$^{-23}$ \\
8.0 & 0.55 & 0.013 & 9.048 & 5.5111$\times$10$^{-23}$ \\
8.5 & 0.83 & 0.009 & 6.264 & 2.6414$\times$10$^{-23}$ \\
9.0 & 1.18 & 0.006 & 4.176 & 1.1740$\times$10$^{-23}$ \\
\end {tabular}
\end{center}
\end{table}

We have attempted to match the EUV spectrum of $\theta$ Hya with a grid
of hot white dwarf$+$ISM model atmospheres, in order to constrain the
possible atmospheric parameters (temperature and surface gravity) of the
degenerate star and the interstellar column densities of HI, HeI and
HeII. Unfortunately, there are no spectral features in this wavelength
region to give us an unambiguous determination of {\it T$_{eff}$} and
{\it log g}. However, by making a range of assumptions to reduce the
number of free parameters in our models, we can place constraints on 
some of the 
the white dwarf's physical parameters. Our method is similar to that
used in the analysis of the white dwarf companion to y Pup (Burleigh and
Barstow 1998). 

Firstly, we assume that the white dwarf has a pure-hydrogen atmosphere.
This is a reasonable assumption to make, since Barstow et~al. (1993)
first showed that for {\it T$_{eff}$}$<$40,000K hot white dwarfs have
an essentially pure-H atmospheric compostion. We can then fit a range of
models, each fixed at a value of the surface gravity {\it log g}. However, 
before we can do this we need to know the normalisation parameter of
each model, which is equivalent to (Radius$_{\it WD}$$/$Distance)$^2$. We
can use the {\it Hipparcos} parallax of 4.34$\pm$0.97
milli-arcsecs., translating to a distance of 39.5$\pm$1.5 parsecs, 
together with the Hamada-Salpeter
zero-temperature mass-radius relation, to give us the radius of the white
dwarf corresponding to each value of the surface gravity (see Table 2). 

\begin{table}
\begin{center}
\caption{WD parameters and interstellar column
densities}
\begin{tabular}{ccccc}
log $g$ & $T_{eff}$ (K) & $N_{HI}$ $\times$10$^{18}$ & $N_{HeI}$ 
& $N_{HeII}$ \\
 & \& 90\% range & \& 90\% range  & $\times$10$^{17}$ & 
$\times$10$^{17}$ \\
7.5 & 25,800 (25,500$-$26,200) & 2.7 (1.1$-$4.9) & 3.1 & 1.1 \\
8.0 & 26,800 (26,400$-$27,100) & 4.6 (3.2$-$6.3) & 5.2 & 1.9 \\
8.5 & 28,500 (28,200$-$28,900) & 6.6 (5.3$-$7.9) & 7.4 & 2.8 \\
9.0 & 30,800 (30,400$-$31,200) & 8.0 (7.0$-$9.1) & 9.0 & 3.3 \\
\end {tabular}
\end{center}
\end{table}

We can also reduce the number of unknown free parameters in the ISM model. 
From EUVE spectroscopy, 
Barstow et al. (1997) measured the line-of-sight interstellar column
densities of HI, HeI and HeII to a number of hot white dwarfs. They found
that the mean H ionisation fraction in the local ISM was 0.35$\pm$0.1,
and the mean He ionisation fraction was 0.27$\pm$0.04. From these
estimates, and assuming a cosmic H/He abundance, we calculate the ratio 
 $N_{HI}$/$N_{HeI}$ in the local ISM$=$8.9, and
$N_{HeI}$/$N_{HeII}$$=$2.7. We can then fix these column density ratios
in our model, leaving us with just two free parameters - temperature and
the HI column density. 
 
The model fits at a range of surface gravities from log $g=$7.5$-$9.0 are
summarized in Table 3. Note that our range of fitted temperatures is in
broad agreement with those of VCT98,
who modelled the EUV and soft X-ray photometric data for $\theta$ Hya on
the assumption that the source was indeed a hot white dwarf.

\section{Discussion}

%\begin{figure*}
%\vspace*{8cm}
%\special{psfile=figure2.ps  
%hscale=110 vscale=70 hoffset=0 voffset=-95}
%\caption{\em IUE short wavelength spectrum of HR3665, shown together with
%a white dwarf model spectrum for the same parameters as Figure 1. This
%diagram shows the Lyman H series of absorption lines, and clearly demonstrates
%that the white dwarf could not be detected in the far-UV by either IUE or HST.}
%\end{figure*}

We have analysed the EUVE spectrum of the B9.5V star $\theta$ Hya which
confirms that it has a hot white dwarf companion, and 
constrains the degenerate star's temperature to lie between $\approx$25,500K
and $\approx$31,000K. This is the second B
star$+$hot white dwarf binary to be spectroscopically identified, following 
y Pup (HR2875), a B5 main sequence star. The  
white dwarf in the $\theta$ Hya system must have evolved from a
progenitor more massive than B9.5V ($\approx$3.4M$_\odot$).  
  
Although EUVE spectra provide us with little
information with which to constrain a white dwarf's surface gravity,
and hence its mass, we can use a theoretical initial-final mass relation
between main sequence stars and white dwarfs, e.g. that of Wood (1992),
to calculate the mass of a white dwarf if the progenitor was only
slightly more massive than $\theta$ Hya:

\vspace{0.35cm}

$M_{WD}$$=$Aexp(B$\times$$M_{MS}$) 

\vspace{0.35cm}

where~A$=$0.49M$_\odot$ and
B$=$0.094M$_\odot$$^{-1}$. 

\vspace{0.35cm}

For $M_{MS}$$=$3.4M$_\odot$, we find $M_{WD}$$=$0.68M$_\odot$.
This would suggest the surface gravity of the white dwarf log g$>$8.0.

Data from {\it Hipparcos}
indicates possible micro-variations in the proper motion of $\theta$ Hya
across the sky, suggesting that the binary period may be $\sim$10 years
or more. Indeed, VCT98 measured
marginal variations in the B star's radial velocity. Clearly, more
measurements at regular intervals in future years might help to pin down
the binary period.

\section{A third B star$+$white dwarf binary in the EUV catalogues?}

EUVE has now spectroscopically identified two B star$+$hot white dwarf
binaries from the EUV all-sky surveys. As mentioned previously in
Section~3, in a survey of X-ray detections of OB stars, Bergh\"ofer,
Schmitt and Cassinelli (1996) found just three B
stars which were soft X-ray sources only: y Pup and $\theta$ Hya, which
have hot white dwarf companions responsible for the EUV and soft X-ray
emissions, and 16 Dra (B9.5V, $=$HD150100, $=$ADS10129C, V$=$5.53). 

16 Dra is one member of a bright resolved triple system (with HD150118,
A1V, and HD150117, B9V). Hipparcos parallaxes confirm all three stars lie
at the same distance, $\approx$120 parsecs. 16 Dra is
also a WFC and EUVE source (RE J1636$+$528), and it is so similar to y
Pup and $\theta$ Hya that we predict it also has
a hot white dwarf companion, most likely unresolved.  
Unfortunately, it is a much fainter EUV
source than either y Pup or $\theta$ Hya, and would require a significant
exposure time to be detected by EUVE's spectrometers ($\sim$400$-$500
ksecs). 

However, we can estimate the approximate temperature of this white dwarf, 
and the neutral hydrogen column density to 16 Dra, using the ROSAT
photometric data points: WFC S1 count rate $=$12$\pm$4 c/ksec,
S2$=$46$\pm$11 c/ksec, and PSPC soft band count rate $=$72$\pm$15 c/ksec.
We adopt a similar method to the analysis of $\theta$ Hya described
earlier, using the Hamada-Salpeter zero-temperature mass-radius relation
and the Hipparcos parallax to constrain the normalisation (equivalent to
($R_{WD}$/$D$)$^2$). Although we cannot constrain the value of the
surface gravity using this method, we find that the white dwarf's
temperature is likely to be between 25,000$-$37,000K, and the neutral
hydrogen column density N$_{HI}$$<$4$\times$10$^{19}$ atoms cm$^{-2}$.

\begin{acknowledgements}

Matt Burleigh and Martin Barstow acknowledge the support of PPARC, UK.
We thank Detlev Koester (Kiel) for the use of his white dwarf model
atmosphere grids. This research has made use of the {\it SIMBAD} database
operated by CDS, Strasbourg, France.

\end{acknowledgements}

\end{document}